\documentclass[aps,twocolumn,pre,showpacs,nofootinbib]{revtex4}
\usepackage{graphicx}

\newcommand{\MCtwo}{Microtechnology and Nanoscience, MC2, 
Chalmers University of Technology, SE-412 96 G{\"o}teborg, Sweden}

\newcommand{\vdW}{{\mbox{\scriptsize vdW-DF}}}
\newcommand{\near}{{\mbox{\scriptsize near}}}
\newcommand{\far}{{\mbox{\scriptsize far}}}
\newcommand{\first}{{\mbox{\scriptsize deform}}}
\newcommand{\last}{{\mbox{\scriptsize gas}}}
\newcommand{\des}{{\mbox{\scriptsize des}}}

\newcommand{\LDA}{{\mbox{\scriptsize LDA}}}
\newcommand{\PBE}{{\mbox{\scriptsize PBE}}}

\newcommand{\nl}{{\mbox{\scriptsize nl}}}

\newcommand{\defi}{{\mbox{\scriptsize def}}}

\begin{document}

\title{Desorption of n-alkanes from graphene: a van der Waals density functional study}

\author{Elisa Londero}\affiliation{\MCtwo}
\author{Emma K. Karlson}\affiliation{\MCtwo}
\author{Marcus Landahl}\affiliation{\MCtwo}
\author{Dimitri Ostrovskii}\affiliation{\MCtwo}
\author{Jonatan D. Rydberg}\affiliation{\MCtwo}
\author{Elsebeth Schr{\"o}der}\thanks{Corresponding author}\email{schroder@chalmers.se}%
\affiliation{\MCtwo}

\date{\today, Submitted to Journal of Physics: Condensed Matter}

%%%%%%%%%%%%%%%%%%%%%%%%%%%%%%%%%%%%%%%%%%%%%%%%%%%%%%%%%%%%%%%%%%%%%%%%%%%%%%%%%%%%%%%%%%%%%%
\begin{abstract} 
A recent study of temperature programmed desorption (TPD) measurements of 
small $n$-alkanes (C$_N$H$_{2N+2}$) from C(0001) deposited on Pt(111) 
shows a linear relationship of the desorption energy with 
increasing $n$-alkane chain length.
%[J.~Chem.\ Phys.\ \textbf{125}, 234308 (2006)].
We here present a van der Waals density functional study of the 
desorption barrier energy of the ten smallest $n$-alkanes ($N=1$ to $10$)  
from graphene. We find linear scaling with $N$, 
including a nonzero intercept with the energy axis, i.e., an offset 
at the extrapolation to $N=0$. 
This calculated offset is quantitatively similar to the results 
of the TPD measurements. From further calculations of the 
polyethylene polymer we offer a suggestion for the origin of the offset. 
\end{abstract}

\maketitle

%%%%%%%%%%%%%%%%%%%%%%%%%%%%%%%%%%%%%%%%%%%%%%%%%%%%%%%%%%%%%%%%%%%%%%%%%%%%%%%%%%%%%%%%%%%%%%

\section{Introduction} 
The increasing use of molecules on graphene and 
graphite surfaces for industrial
applications calls for an improved atomic-scale understanding of the 
adsorption/desorption structure and process. 
The $n$-alkanes are linear chains of hydrocarbons, short versions of 
the polyethylene (PE) polymer.
Using temperature-programmed desorption (TPD) 
Tait et al.\ (Ref.~\onlinecite{tait2006}) measured the
desorption energy and desorption rate preexponential factor 
of $n$-alkanes on graphene 
deposited on a Pt(111) substrate. The $n$-alkanes measured
were short, with the number of C atoms $N\leq 10$.
The desorption of $n$-alkanes from graphite surfaces was also measured 
by Paserba and Gellman \cite{paserba2001,gellman2002,paserba2001b}, 
and from various other surfaces  
by a number of other groups \cite{tait2006refs}, the surface materials including
metals (Ag, Au, Cu, Pt, Ru), oxides (Al$_2$O$_3$, MgO),
and semiconductors (Si).

In most of the alkane desorption measurements the desorption energy was found to scale 
linearly with $N$ for the short $n$-alkanes, but with a non-zero 
intercept with the axis of the desorption energy.  
The value found for this offset at $N=0$  
was sometimes found to be unphysically large, several times larger than the scaling coefficient.
In another study Tait et al. \cite{tait2005-707,tait2005}  
analyzed their own data for $n$-alkane on MgO(100) 
desorption. They allowed the desorption prefactor to vary with 
chain length and found the desorption energy offset to be non-vanishing but small, 
of the size of or smaller than the scaling coefficient.
When the same group of authors analyzed their
data of $n$-alkanes on Pt(111) and on graphene (and also re-analyzed data from a
number of studies by other groups for some of the above-mentioned surfaces) they  
found similar non-vanishing but small offsets for those desorption systems.   
 
In this paper we use the first-principles van der Waals (vdW) 
density-functional method \cite{Dion,Thonhauser}, vdW-DF, to  
determine the $n$-alkane adsorption energy on graphene at low 
coverages for short alkane chains ($N\leq 10$). This adsorption energy
can be compared with the experimentally determined desorption
barrier energy values. 
As in the experimental 
studies in Refs.~\onlinecite{tait2006,tait2005-707}, and \onlinecite{tait2005}
we find a close-to-linear growth in 
adsorption energy $E_a$ with chain length $N$, with a non-vanishing but small offset 
when extrapolated to $N=0$ 
\begin{equation}
E_a= 7.23 N + 6.44 \mbox{ [kJ/mol].}  
\end{equation} 
Here and below we use the term adsorption energy ($E_a$) for the energy found 
in our theory calculations. This corresponds to the desorption energy 
$E_0$ of isolated alkane molecules on graphene.
By $E_d$ we denote the experimental desorption energy, or desorption barrier, 
of an alkane molecule from partly covered graphene. In parts of the literature $E_d$
is instead denoted $\Delta E_\des^\ddagger$. 

Contrary to analysis of the experiments, our calculations of the adsorption/desorption
energy do not involve an assessment of the  
desorption prefactor. Our values of $E_a$ are simply found from 
the differences in total energies of the system in the adsorbed and the desorbed states. 

The outline of the paper is as follows. In Section II we describe the method of computation,
including convergence tests of sensitive computational parameters.
Section III presents our results and discussions, including a discussion of
the definition of a monolayer (ML) of coverage of $n$-alkanes on graphene.
Section IV contains our summary. 
%%%%%%%%%%%%%%%%%%%%%%%%%%%%%%%%%%%%%%%%%%%%%%%%%%%%%%%%%%%%%%%%%%%%%%%%%%%%%%%%%%%%%%%%%%%%%%

\section{Method of computation}
The $n$-alkanes are linear, saturated hydrocarbon chains absent of branches, 
with the general formula C$_N$H$_{2N+2}$, $N > 0$. 
Very long such chains (in principle infinitely long) are known as the PE polymer. 
In this paper we analyze the adsorption on graphene of the ten smallest
$n$-alkanes ($1\leq N \leq 10$), of H$_2$, and of PE, all in the stretched form,
which is the trans conformation.

Our interest in the alkane desorption was sparked by the TPD experiments 
of Tait et al.\ \cite{tait2006} and their analysis leading to the 
experimentally determined desorption energy.
We determine the adsorption energy 
by use of first-principles density functional theory (DFT), employing 
the method vdW-DF \cite{Dion,Thonhauser} as detailed in several other
publications \cite{PAHgraphite,PAHgg,adenine} but here with the vdW interaction
treated fully selfconsistently \cite{Thonhauser}.
We calculate the total energies of the adsorption system 
using the DFT program \textsc{gpaw} \cite{gpaw} with vdW-DF \cite{Dion,Thonhauser}
in a Fast-Fourier-implementation \cite{soler}.
 
\begin{figure}[bt]
\begin{center}
\includegraphics[width=0.4\textwidth]{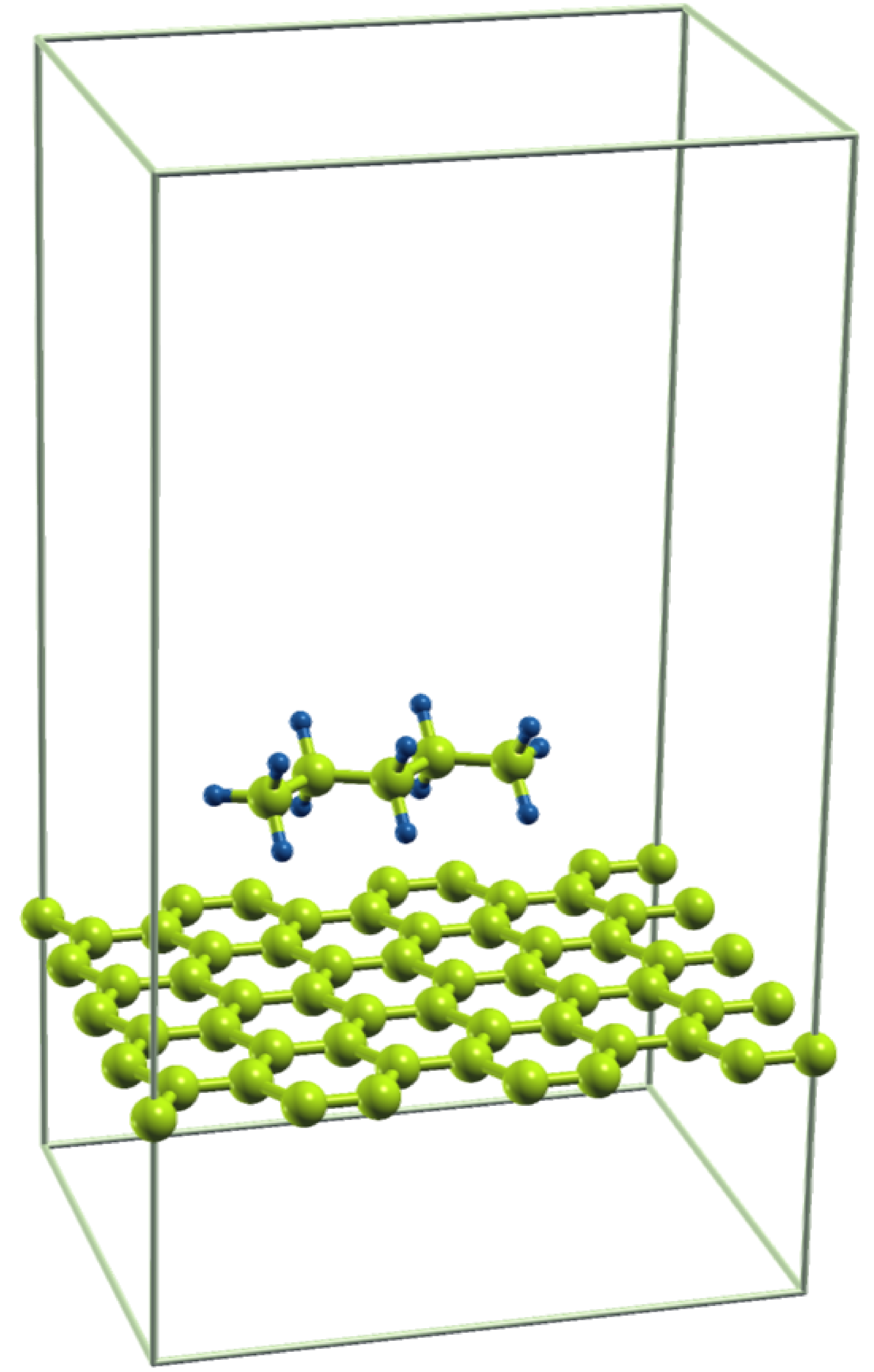}
\caption{Illustration of $n$-pentane ($N=5$) adsorbed on graphene. One 
unit cell is shown including a repetition 
of the graphene carbon atoms that are positioned on the unit cell boundary. 
C atoms are represented by large spheres and H atoms by
small spheres.
Figure created using \textsc{XCrySDen} \protect\cite{XCrySDen}.
}
\label{fig:nalkanes}
\end{center}
\end{figure}

Figure \ref{fig:nalkanes} illustrates the adsorbed $n$-pentane molecule
on graphene, and the unit cell used in our calculations for $N=5$.
The lateral sizes of all the unit cells used are listed
in Table \ref{tab:desorption}.

For each of the adsorbed alkane molecules we determine the optimal positions
of the atoms by minimizing the Hellmann-Feynman forces.
These are derived from gradients in the 
self-consistently determined electron density.
This optimization also adjusts the intramolecular bond lengths to the 
most favorable value in the adsorption state. After this optimization  
we obtain the total energy of the adsorbate-graphene system, $E^\vdW_\near$.

The \textsc{gpaw} code is an all-electron DFT code based on 
projector augmented waves \cite{Blochl} (PAW)
and using finite differences.
In several of our previous vdW-DF applications  we used self-consistent 
calculations with the generalized gradient approximation (GGA) to determine 
the electron density and part of the total energy, 
followed by non-selfconsistent
calculations to determine the total energy within vdW-DF. This allowed us
to focus on either the GGA need for accuracy in choice of computational
parameters and methods (for example the need of a fixed amount of vacuum in 
the total system \cite{benzene,londero2010,londero2011})
or the vdW-DF need for accuracy in other parameters and methods (for example 
the need for fixating the local electron density grid \cite{PAHgg,Ziambaras,NTgg}). 
When carrying out selfconsistent calculations for vdW-DF all of these 
requirements must be met in all calculations. Below the most important 
of these choices are described. 

We model the adsorption system by means of an orthorhombic unit cell, 
periodically repeated in all directions.
The unit cell contains sufficient space in the plane of graphene
for the repeated images of the $n$-alkane molecules to interact
only little. 
As explained below, we make sure to explicitly
subtract the small lateral interaction of the periodic images of the
molecules from our results \cite{adenine,slidingrings}.
In the direction perpendicular to graphene the introduction 
of $\sim 19$ {\AA} of vacuum above
the adsorbed alkane ensures that no interaction across unit cell boundaries
takes place.

For the smallest of the alkanes, methane ($N=1$), we use
a unit cell of size $3\sqrt{3}\,a_g\times 3\,a_g \times 23.0$ {\AA}
and for the largest alkane considered here, $n$-decane ($N=10$), we use a
$5\sqrt{3}\,a_g\times 4\,a_g \times 23.0$ {\AA} unit cell.
Here $a_g= \sqrt{3}\, a_0$, with $a_0=1.43$ {\AA}, is the clean graphene lattice
constant as found in our calculations by relaxing the lateral size of the unit cell.
The unit cell sizes used for the other
calculations ($N=2, \ldots , 9$) are listed in Table~\ref{tab:desorption}.

We choose the 
real-space grid for representing the wavefunctions 
in the PAW procedure to have a distance less than 
0.11~{\AA} between nearest-neighbor (nn) grid points. The (valence) electron density
is represented on the same grid with additional grid points at half the nn distance,  
values obtained from interpolation of the electron density and 
addition of compensation charges.
The choice of these relatively dense (wavefunction and electron density) grids is
important for the quality of the evaluation of the nonlocal 
correlation contribution \cite{Ziambaras}.

The Brillouin zone of the unit cell is sampled according to the Monkhorst-Pack scheme
by means of a 2$\times$2$\times$1 $k$-point sampling.
Increasing the $k$-point sampling to 
4$\times$4$\times$1 changes $E_a$ less than 0.7 meV per molecule (0.07 kJ/mol). 
We further make sure that the calculation is accurately converged with 
respect to total energies. We impose a convergence 
threshold such that the total energy changes 
less than 0.1 meV per unit cell, or less than approximately 10$^{-6}$ eV 
per atom in the unit cell, in the last three iterations.
This convergence threshold is several orders of magnitude smaller than 
the default settings of \textsc{gpaw}.

\begin{table*}
\caption{\label{tab:desorption} Adsorption (desorption) energies from theory ($E_a$) and
experiment ($E_d$ and $E_0$),
center-of-mass distance from
graphene $d_{cm}$, area $A$ of one alkane molecule in a full monolayer, unit cell used in calculations,
and corresponding estimated coverages $\theta$,
for the small $n$-alkanes (C$_N$H$_{2N+2}$, $N=1-10$). In our calculations we use orthogonal unit cells and
a graphite lattice vector $a_g=\sqrt{3} \,a_0$ with $a_0=1.43$ {\AA}.
The experimental values of $E_0$ --- corresponding to the limit of 0 ML 
coverage and no defect sites --- is found from 
Eq.\ (\protect\ref{eq:experiment}) with the use of parameters given in 
Table IV of Ref.~\protect\onlinecite{tait2006}. The coverage for PE is 
found from an estimate of the PE-PE interaction distance in the PE 
crystal, as described in the text.}

\begin{tabular}{lccccccccc}
 \hline\hline
              &   & \multicolumn{5}{c}{This work}                         &\multicolumn{3}{c}{Experiments}  \\
              &   & \multicolumn{5}{c}{\line(1,0){130}}                   &\multicolumn{3}{c}{\line(1,0){150}}\\
              &$N$& Unitcell            &$\theta$&$d_{cm}$&\multicolumn{2}{c}{$E_a$}&$A$&$E_d(0.5\mbox{ ML})$&$E_0$ \\
              &   &                     & [ML]   &[\AA]   &[kJ/mol]& [eV]  &[{\AA}$^2$]&[kJ/mol]&[kJ/mol] \\
  \hline
  H$_2$       & 0 & $2\sqrt{3}\times 3$ &        & 3.39   &  6.8   & 0.070 &           &        &      \\
  methane     & 1 & $3\sqrt{3}\times 3$ & 0.16   & 3.64   & 14.6   & 0.152 & 15$^a$    &  14.1  & 13.6 \\
  ethane      & 2 & $3\sqrt{3}\times 3$ & 0.22   & 3.80   & 20.9   & 0.216 & 20.9$^b$  &  24.6  & 23.8 \\
  propane     & 3 & $3\sqrt{3}\times 3$ & 0.28   & 3.90   & 27.7   & 0.288 & 27$^a$    &  32.1  & 30.6 \\
  $n-$butane  & 4 & $3\sqrt{3}\times 4$ & 0.26   & 3.97   & 34.6   & 0.358 & 32.7$^c$  &  40.8  & 38.9 \\
  $n-$pentane & 5 & $3\sqrt{3}\times 4$ & 0.31   & 3.86   & 42.8   & 0.443 & 39$^a$    &        &      \\
  $n-$hexane  & 6 & $4\sqrt{3}\times 4$ & 0.26   & 3.96   & 49.6   & 0.514 & 45.6$^d$,
                                                                             44.8$^e$  &  63.0  & 60.3 \\
  $n-$heptane & 7 & $4\sqrt{3}\times 4$ & 0.30   & 3.90   & 57.7   & 0.598 & 51.6$^f$  &        &      \\
  $n-$octane  & 8 & $5\sqrt{3}\times 4$ & 0.27   & 3.89   & 65.5   & 0.679 & 57.2$^f$,
                                                                             57.7$^d$,
                                                                             56.2$^e$  &  72.6  & 71.0 \\
  $n-$nonane  & 9 & $5\sqrt{3}\times 4$ & 0.30   & 3.87   & 73.0   & 0.757 & 63.5$^f$  &        &      \\
  $n-$decane  &10 & $5\sqrt{3}\times 4$ & 0.32   & 3.86   & 80.3   & 0.832 & 69.0$^f$,
                                                                             69.7$^d$,
                                                                             68.9$^e$  &  91.4  & 84.5 \\
  polyethylene&(1)& $5\sqrt{3}\times 1$ & 0.21   & 3.83   & 7.2$^g$& 0.074$^g$&        &        &      \\
\hline
\multicolumn{10}{l}{${}^a$Linear interpolation of the experimental data available for other values of
$N$, $A(N)\approx 9+6N$ {\AA}$^2$.}\\
\multicolumn{10}{l}{${}^b$Neutron diffraction data at submonolayer coverage, Ref.~\protect\onlinecite{hansen1984}.}\\
\multicolumn{10}{l}{${}^c$Neutron diffraction data at 11 K, Ref.~\protect\onlinecite{herwig1994}.}\\
\multicolumn{10}{l}{${}^d$X-ray diffraction data at submonolayer coverage, Ref.~\protect\onlinecite{arnold2002}.}\\
\multicolumn{10}{l}{${}^e$Neutron diffraction data at submonolayer coverage, Ref.~\protect\onlinecite{arnold2002}.}\\
\multicolumn{10}{l}{${}^f$X-ray diffraction data, Ref.~\protect\onlinecite{inaba2002}.}\\
\multicolumn{10}{l}{${}^g E_a$ per C atom in PE. Each unit cell has two units of CH$_2$.}
    \end{tabular}
\end{table*}

We determine the adsorption energy  $E_a$
as the difference in total energy between a system with an alkane adsorbed 
in the optimal geometry and a system with the alkane moved far away from graphene.
It is well known that in the vacuum region of the system small spurious 
exchange energy contributions add to the 
total-energy \cite{benzene,londero2010,londero2011}. 
To cancel these contributions we use the same 
unit cell for the calculation of the adsorbed state and for the desorbed state.
We make sure 
that the height of the unit cell allows the fragments (graphene and alkane molecule) to be
far apart within the unit cell. With unit cell height 23 {\AA} the maximum 
possible separation is $\sim 11$ {\AA}, sufficient for the alkane to count as desorbed.

The correlation energy $E_c$ in the total energy 
for the vdW-DF functional is split \cite{IJQC} into a nearly-local part $E_c^0$ and a part that
includes the most nonlocal interactions $E_c^\nl$,
\begin{equation}\label{eq:1}
E_c = E_c^0+E_c^\nl \,.
\end{equation}
In a homogeneous system the term $E^0_c$ is the correlation $E^\LDA_c$ obtained from the local
density approximation (LDA),
and in general \cite{Dion} we approximate $E^0_c$ by $E^\LDA_c$.
The term $E_c^\nl$ vanishes for a homogeneous system.
It describes the dispersion interaction.
The form of $E_c^\nl$ is derived in Ref.~\onlinecite{Dion}.
It is a truly nonlocal functional  
\begin{equation}\label{eq:2}
E_c^\nl[n] =
\frac{1}{2}\int\int d\mathbf{r}\,d\mathbf{r}'\,n(\mathbf{r})\phi(\mathbf{r},\mathbf{r}')n(\mathbf{r}')
\end{equation}
given by a kernel $\phi$ which is explicitly stated in Ref.~\onlinecite{Dion}.

$E_c^\nl$ exhibits a small sensitivity  
to changes in the local real space grid, for example 
translations of the nuclei positions by a non-integer
number of real-space grid points \cite{PAHgg,Ziambaras,NTgg}. 
The procedure to ensure high-quality results that include contributions from 
changes in the intra-molecular bonds 
is described in more detail in Refs.\ \onlinecite{h2cu} and
\onlinecite{phenol}, and is reviewed here.

We derive the adsorption energy in three steps. 
First we carry out a vdW-DF calculation of the adsorbed molecule, yielding 
the total energy $E^\vdW_\near$. Next we carry out a vdW-DF calculation of the molecule moved 
off graphene rigidly, which gives us the total energy $E^\vdW_\far$.  
Finally, we let the molecule relax into the gas phase structure. 
{}From the last step  we obtain values of the total energy in
the first (deformed for adsorption) and in the last (gas phase) situation,
$E^\PBE_\first$ and $E^\PBE_\last$ respectively. In this last relaxation study step we
use the Perdew-Burke-Ernzerhof \cite{PBE} (PBE) variant of GGA. 
The adsorption energy $E_a$, shown in Table \ref{tab:desorption}, is then found from
\begin{equation}
-E_a=E^\vdW_\near-E^\vdW_\far+E^\PBE_\first-E^\PBE_\last.
\label{eq:Ed}
\end{equation}
  
For the calculation of $E^\vdW_\near$ we determine the optimal 
position of the alkane molecule adsorbed on graphene.
The Hellmann-Feynman forces on the alkane atoms are minimized. 
All graphene atoms are fixed in space in this and all other calculations.  

For the calculation of  $E^\vdW_\far$, 
we move the alkane molecule (rigidly) off of the graphene sheet by
translating the molecule by 75 grid points, corresponding to adding $\sim 8.13$ {\AA}
to the graphene-alkane distance.  
This yields an alkane-graphene separation of approximately 11-12 {\AA} 
both between fragments within the unit cell and fragments in the vertically 
repeated images. By translating the alkane an integer number of grid points 
the nuclei locally maintain the same positions 
with respect to the grid, thus avoiding any possible effects of describing
the electron density differently on a shifted grid \cite{Ziambaras,NTgg}.
This intermediate reference energy  $E^\vdW_\far$ is then subtracted
from  $E^\vdW_\near$. 
We use the same unit cell size, number of real space grid points, and number 
of $k$-points as for the adsorption calculation.

The use of  
an intermediate reference energy subtracts not only intra-molecular 
contributions (the alkane molecule is identical
in the two calculations) but also any direct alkane-alkane interaction across unit cell
boundaries. This is usually a small contribution in our calculations.

The last set of calculations finds the energy gain 
$E^\PBE_\first-E^\PBE_\last$ of relaxing the isolated alkane molecule 
from the slightly deformed structure of the adsorbed 
state. 
This energy gain is not expected to have any long-range component.
The $E_c^\nl$ term of vdW-DF depends slightly on the 
density grid points positions with respect to the molecule, and we 
therefore, as a matter of principle,  
choose to perform this part of the calculation using instead the GGA variant 
PBE \cite{PBE}.
 
It should be noted, though, that in this study it does not
numerically matter to the energetics whether we use PBE or vdW-DF in this 
last relaxation-study step.

\begin{figure}
\begin{center}
\includegraphics[width=0.47\textwidth]{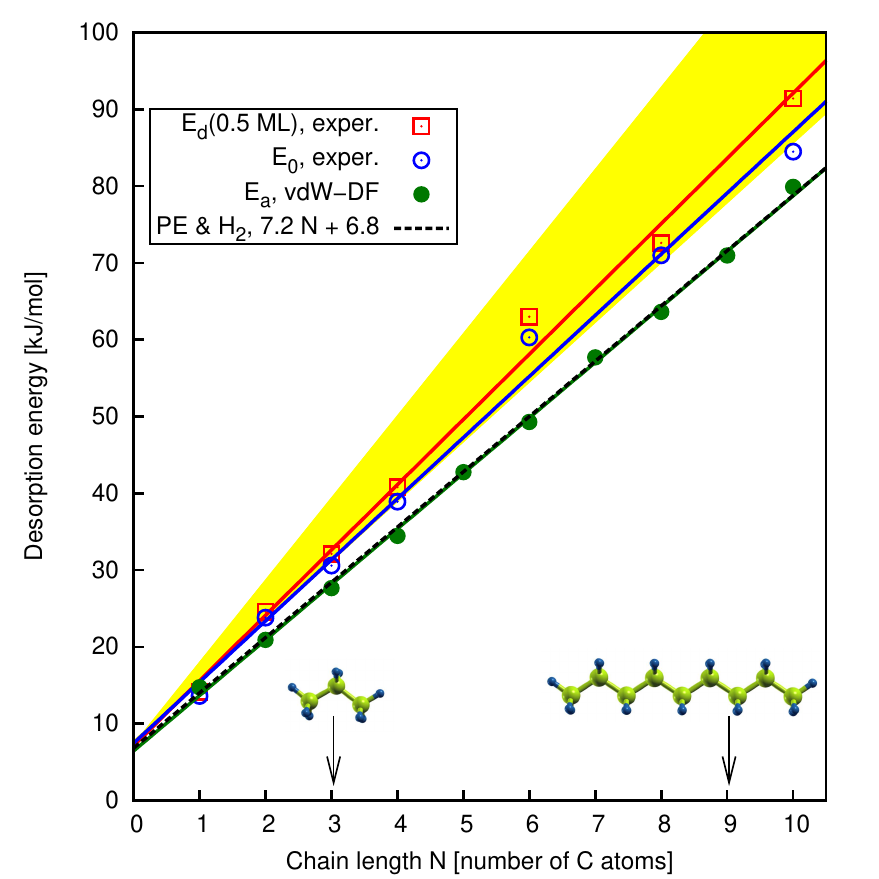}
\caption{\label{fig:linear}Desorption energy as a function of the length 
of the $n$-alkane chain. Solid points are our results, open points are 
from the TPD measurements by Tait et al.\ \protect\cite{tait2006}.
Linear regression lines for the three sets of data points are also shown, 
including the extrapolations to $N=0$. The dashed line has the slope from
our calculations of the adsorption of PE, with the H$_2$ adsorption energy
added to represent the ends of the alkanes. The shaded area encompasses
the estimated (by Ref.~\protect\onlinecite{tait2006}) errorbars of the 
0.5 ML experimental data.
} 
\end{center}
\end{figure}

%%%%%%%%%%%%%%%%%%%%%%%%%%%%%%%%%%%%%%%%%%%%%%%%%%%%%%%%%%%%%%%%%%%%%%%%%%%%%%%%%%%%%%%%%%%%%% 
\section{Results and discussion}
The alkane molecules deform very little upon adsorption,
compared to their gas phase structure.
In the gas phase, calculated with vdW-DF, we find for $n$-pentane 
the average C-C bond length 1.541 {\AA}.  
The C-C distance varies slightly along the carbon chain, with the
smallest values towards the ends and the largest values around
the center of the chain, but the difference only amounts to 0.001 {\AA}.
For $n$-nonane the C-C bond lengths along the chain differ by 0.002 {\AA},
again with the largest values around the center of the chain.
We find similar bond lengths and bond length variations for the other alkanes.

Values of the average C-C bond length found by experiment \cite{nist} for
$n$-alkanes with $N=2$ to 7 are in the range 1.526 {\AA} ($n$-propane) to
1.536 {\AA} (ethane). 
Our results for the bond lengths thus deviate less than 1\% from 
experiment.\footnote{When comparing the values of the 
bond lengths with experiment it should be kept in 
mind that many exchange-correlation approximations,
like the vdW-DF and also many of the GGA approximations,
find covalent bond lengths that can deviate up to a few percent from the 
experimental values.}

When an alkane molecule is adsorbed on graphene we find that 
the bond lengths are only slightly affected. 
For pentane the adsorbed molecule (Fig.\ \ref{fig:nalkanes}) has an
average C-C bond length 1.543 {\AA}, a change from the gas phase by 
0.002 {\AA} (or 0.1\%), and the bond lengths are still larger towards 
the middle of the chain (1.544 {\AA}) compared to the bonds at the 
ends of the chain (1.542 {\AA}). For the other alkanes we find the 
average C-C bond length in the range 1.540 {\AA} (ethane) to 
1.544 {\AA} (octane and nonane), and
for all alkanes the bonds towards the middle of the chains are longer
than those at the ends.

Thus, structural changes caused by the adsorption are very small. 
Energetically, the changes are also very small: using PBE for the 
reasons stated in the
previous section we find the difference in total energy 
$E^\PBE_\first-E^\PBE_\last$
approximately 2 meV (0.2 kJ/mol) per C-C bond for all the alkanes 
studied here. 

All calculations of the adsorbed alkanes presented above are for
an orientation with the alkane carbon skeleton parallel to graphene.
We started out the process of optimizing the atomic positions 
with the carbon skeleton parallel to graphene, and all calculations
reached an energetic minimum when parallel to graphene. 
To check if this is also the global minimum we further 
calculated the total energy for alkane molecules that initially 
were oriented with their carbon skeleton perpendicular
to graphene, i.e., rotated 90$^\circ$ around their axis.
For these, we also find a local minimum, but with a total energy
larger (less favorable) than for the configuration with
the backbone parallel to graphene. For example for pentane we find
that the loss of total energy going from the perpendicular orientation
to the parallel orientation is 48 meV (4.6 kJ/mol). 

Table \ref{tab:desorption} lists the adsorption (desorption) energies obtained 
from theory by us and through TPD measurements by Tait et al.\ \cite{tait2006}.
As shown in Figure \ref{fig:linear} the calculated adsorption energy
values grow linearly with the size of the alkane molecule, $N$,
with an off-set comparable to that from experiments.
Although the coverage of adsorbed alkanes in our calculation is 0.2--0.3 ML
(with full coverage defined as described further below) the subtraction procedure
involving the two first terms in (\ref{eq:Ed}) ensures that all direct alkane-alkane
interactions are eliminated. Thus our results should be compared with 
experimental results for single alkane molecules desorbed from otherwise
clean and defectless graphene (coverage 0 ML), $E_0$.

In Ref.~\onlinecite{tait2006}, the model used for describing the
desorption energy $E_d$ as a function of the coverage $\theta$ and the number
of alkane carbon atoms $N$ is
\begin{equation}
E_d(\theta,N)= E_0(N) +\gamma(N)\theta 
+ E_\defi(N) \mathrm{exp}\left({-\frac{\theta}{\theta_{\mathrm{def}} (N)}}\right)\,.
\label{eq:experiment}
\end{equation}
The $E_0(N)$ is the contribution from a defectless surface (here: 
graphene) in the absence of adsorbate-adsorbate interactions. The term 
$\gamma(N)\theta$ accounts for the increase in desorption energy due to 
the interaction with other adsorbates on the surface, and the third 
term describes the effect of defects in the surface.
The model is introduced in Ref.\ \onlinecite{tait2005-707} for
$n$-butane on MgO(100). For general (small) $n$-alkanes on graphite 
the parameters $\gamma$, $E_\defi$, and $\theta_{\mathrm{def}}$ 
are given in Table IV of Ref.~\onlinecite{tait2006}.
In our calculations graphene is defectless and there are no lateral 
interactions between molecules. Therefore our adsorption  
energies $E_a$ should be compared with the experimental quantity $E_0$,
listed in Table~\ref{tab:desorption}.

In Figure \ref{fig:linear} are shown the experimental results $E_d$ 
at $\theta=0.5$ ML, with error estimates within the
shaded area (from Table III of Ref.\ \onlinecite{tait2006}),  and $E_0$
at zero coverage, together with our adsorption energies $E_a$.
Comparing $E_a$ with $E_0$ we find that  
our theory adsorption energies deviate somewhat from the experimental 
results, with values from theory about 10\% smaller
than $E_0$. Hexane is an exception: our vdW-DF value $E_a$ is 
18\% smaller than $E_0$. However,  
for hexane the experiment deviates from the 
linear growth with $N$ whereas our theory result does not. 

The solid linear curves in Figure \ref{fig:linear} are the linear regression curves for  
$E_d$, $E_0$, and $E_a$. 
The experimental curves are described by 
$E_d= 8.50 N + 7.11$  
(Ref.~\onlinecite{tait2006}) and $E_0= 7.96 N + 7.46$, and we find  
for the theory results the relationship $E_a= 7.23 N + 6.44$, 
all given in units of kJ/mol.

The shaded area in Figure \ref{fig:linear} shows the estimated errors 
in the values of $E_d$ as provided by Ref.~\onlinecite{tait2006}; 
it is reasonable to expect 
similar error estimates on the experimental $E_0$ values but these are not
available to us.

\begin{figure}
\begin{center}
\includegraphics[width=0.47\textwidth]{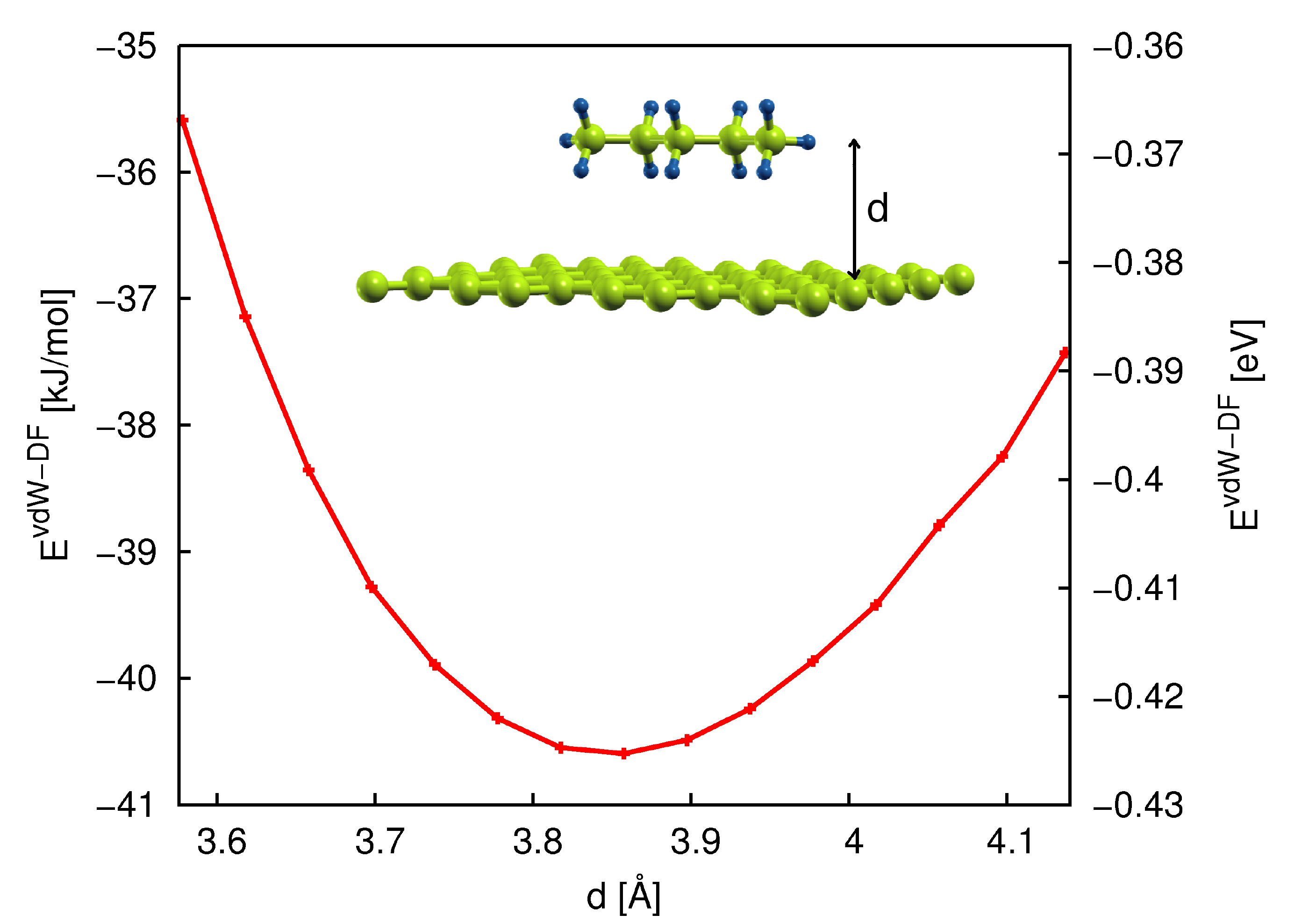}
\caption{\label{fig:curve} Potential energy for $n$-pentane on graphene.
The potential is calculated like $-E_a$ in (\protect\ref{eq:Ed}) but
with the center of mass of pentane fixed at the distances $d$ from 
graphene. 
The minimum of the curve corresponds to the adsorption energy $E_a$ and 
distance $d_{cm}$.
}
\end{center}
\end{figure}

In Figure \ref{fig:curve} we show the potential energy curve 
for $n$-pentane as pentane is moved away from graphene, 
obtained with the vdW-DF functional.
The points of the curve are obtained as described by (\ref{eq:Ed}),
with three sets of calculations, however
with the center of mass of the molecule kept fixed at 
the distance $d$ above graphene.
All internal atomic positions of pentane are allowed to relax. 

Figure \ref{fig:curve} shows that the potential is
shallow around the adsorption position. For example, moving the center
of mass of pentane 0.1 {\AA} towards or away from graphene results in
an energy increase of only about 1 kJ/mol, or 10 meV.  
For our calculations we also report in Table \ref{tab:desorption}
the distance $d_{cm}$ between the center of
mass of the molecule and the graphene sheet at the adsorption distance,
which ranges from 3.64 {\AA} (methane) to 3.96 {\AA} (hexane).
Because of the shallow minimum there is
some uncertaincy in determining $d_{cm}$. 

Although we subtract the direct alkane-alkane interactions from the adsorption
energy it is in principle possible that small indirect alkane-alkane interactions remain 
if the molecules are placed too close, e.g., closer than in a monolayer coverage. 
Such indirect interactions 
would affect the atomic positions of alkane from nearby alkanes.
We must therefore make sure that the alkanes are sufficiently far apart 
that such indirect interactions are negligible. To quantify this, we determined the 
coverage of alkanes in our calculations in units of ML, for each of the alkanes as 
well as for H$_2$ and PE.  

We describe the coverage of molecules on graphene by fractions $\theta$ 
of a molecular ML. One ML is a one-molecule thick coating of a surface, 
as found by experiments. In order to find $\theta$ for our calculations 
we need to know for a full ML how many molecules cover a specified area
of graphene, or how large an area $A$ does one molecule cover, in average.
Couto et al.\ \cite{couto} found by means of STM that for various 
$n$-alkanes adsorbed on graphite the coating layer is highly ordered. 
The ordering at such high coverage is affected both by the
adsorbate-adsorbate and the adsorbate-substrate interactions.
Disordered arrangement is only activated above a critical temperature.
For most of the molecules we find the definition of 1 ML from experiments 
reported in the 
literature \cite{hansen1984,herwig1994,arnold2002,inaba2002}, but for
methane, propane, and pentane we use an estimate based on a linear 
interpolation of the experimental data available for other values of
$N$, $A(N)\approx 9+6N$ {\AA}$^2$. From the values of $A(N)$ and the 
sizes of our unit cells we calculate the coverages $\theta(N)$ used 
in our calculations.

For PE we estimate $\theta$ from the PE-PE interaction distance in the 
PE crystal. In an earlier study \cite{PEgg} one of us
identified the optimal PE crystal structure within a vdW-DF characterization. 
{}From that study we estimate the optimal
centerline-to-centerline distance between the polymers to be 4.5 {\AA}.
The separation of the PE polymers used in the present study is
$5\sqrt{3} \,a_g\approx 21$ {\AA}. This gives a coverage for PE 
$\theta_{\mbox{\scriptsize PE}} \approx 4.5$ {\AA}$/21$ {\AA} $\approx 0.21$.

As summarized in Table~\ref{tab:desorption}, the coverages used in our study 
are always less than a third of a ML, with values $\theta=0.16 - 0.32$, 
sufficiently sparse for indirect interaction effects to
be neglected.

Almost all the available experimental results on alkane desorption from various
surfaces derive from TPD measurements measuring the desorption rate $r$. 
In order to extract the desorption energy from $r$ the preexponential desorption rate $\nu$
was earlier often assumed to have the value $10^{13}$ s$^{-1}$. This value is accepted as a reasonable 
value for first-order processes in surface physics of atoms and is derived from 
traditional transition state theory.
However, the more complex processes of molecular desorption are not necessarily 
as well described by that particular value, nor more generally by a value that is constant for all 
$n$-alkanes. 

The desorption rate $r$ may be described by the Polanyi-Wigner equation
\begin{equation}
r(\theta,T)=-\frac{d\theta}{dt}(\theta,T)=\nu(\theta,T)\theta^n e^{-E_d(\theta)/k_B T}
\end{equation}
for $n$th order desorption, here $n=1$.
Assuming a constant value of $\nu$ for the small ($N<12$) $n$-alkanes 
the TPD desorption rates give linear growth in $E_d$ with number of 
alkane segments $N$ but with a very large offset \cite{tait2006refs} at 
$N=0$. The offset is much larger than the segmental increment in $E_d$. 
Lei et al.\ speculated \cite{lei}, and Tait et al.\ showed from analysis 
of TPD experiments \cite{tait2005,tait2006}, that $\nu$ takes other and 
varying values in alkane desorption. This was shown for various surfaces 
like graphite, Pt(111) and MgO(100). By treating $\nu$ as a fitting 
parameter along with $E_d$, modified and varying values of $\nu$ are 
found. Such analysis leads to a more modest value of the offset of 
$E_d$ at $N=0$, at the size of or smaller than
the segmental increment in $E_d$ \cite{tait2005,tait2006}. 

In particular, it was found \cite{tait2006} that on graphite $\nu$ 
varies from $10^{13.0}$ s$^{-1}$ for methane to 
$10^{17.8}$ s$^{-1}$ for $n$-decane. Thus the small molecules have
a desorption prefactor similar to that from theory for atoms, whereas 
the prefactors for the larger molecules deviate strongly from this.
Taking these variations into account 
the desorption energy offset at $N=0$ is reduced to 7.11 kJ/mol, with
a segmental increment in $E_d$ of 8.50 kJ/mol. Similar results were 
obtained for small-$N$ $n$-alkane desorption from Pt(111) and MgO(100). 

It thus seems that the previously published large values of the $N=0$ 
offsets can mostly be explained \cite{tait2006,tait2005-707,tait2005} 
as an effect of not allowing $\nu$ to vary for the small $n$-alkanes.
Nevertheless, an offset of a smaller size does remain even in the 
re-analyzed data. 

In the literature the origin of the offset has been debated 
\cite{tait2006,tait2006refs,lei}.
Even though the values of the offsets may be reduced as discussed above, 
also the remaining offset begs an explanation.
Lei et al.\ summarize the discussions by listing a number of suggested reasons: 
(i) the different binding to the surface of the methyl end groups 
(-CH$_3$) compared to the methylene segments (-CH$_2$-);
(ii) the effect from the chain length dependence on the polarizability 
of the alkanes;
(iii) the effect of needing different temperatures for the various 
alkanes for measuring $r$;
(iv) possibly the desorption process cannot be described as a first-order 
process, e.g., if the alkanes adsorb in islands or other structures;
(v) possible lattice mismatch of the alkanes with the surface; and finally 
(vi) chain length dependence of $\nu$. 
The latter suggestion reduces the offset to a more modest value, 
as discussed above.
 
Without going into details of all of the above-mentioned suggestions we 
note that our calculations are in a sense more direct than the 
desorption energies derived from the TPD measurements.
In our calculations the preexponential factor $\nu$ is not involved, 
temperature variation is not an issue, and we do not let the alkanes 
adsorb in islands.  Our results are
in agreement with the results presented in Ref.~\onlinecite{tait2006} where
the approach of a variable desorption prefactor was used. In particular,
our theoretically calculated value of the offset agrees very well with 
that obtained by Tait et al. Here we present a simple model study to 
discuss the suggestion (i) of end-group effects.

Our calculated adsorption energy for PE corresponds very well with our 
similarly calculated adsorption energy per segment of the alkanes 
(when neglecting the offset). PE is similar to the alkanes,
but it does not (at least not ideally) include methyl end groups.
In our calculations we describe PE adsorbed on graphene by periodically 
repeating two CH$_2$ units, thus explicitly avoiding end groups. 
We find (Table \ref{tab:desorption}) the adsorption energy per methylene 
unit in PE, 7.2 kJ/mol, which corresponds very well to the energy 
7.4 kJ/mol we find per (methylene or methyl) unit for small $n$-alkanes.  

It is natural to expect that the two extra H atoms attached to the ends of the 
alkane molecules (in the methyl groups) also contribute to the adhesion, thus
affecting the offset in the adsorption energy.
We present a calculation of a H$_2$ molecule adsorbed on graphene to 
test a hypothesis of simple additivity of
$N$ methylene segments (-CH$_2$-) and two additional H atoms. 
This is thus an even simpler model for $n$-alkane than adding
methyl to the ends of a string of methylene segments.

Our calculation of H$_2$ on graphene yields the adsorption energy 6.8 kJ/mol.
In the calculated curve for $E_a$ the offset is 6.44 kJ/mol. 
Our results for PE and 
H$_2$ fit nicely to this simple additivity model.
The curve with slope derived from PE adsorption and
offset derived from H$_2$ adsorption is plotted in Figure \ref{fig:linear} (dashed line).
 
Arguments raised against the end-group explanation have been that 
experiments \cite{lei} for cyclic alkanes on Cu(111) and Pt(111) 
also show an offset for extrapolation to $N$ even though the
cyclic alkanes do not have any end groups.
However, those results were extracted using fixed values of $\nu$ and 
yield large offsets (36 kJ/mol for Pt, 19 kJ/mol for Cu) both for the 
cyclic alkanes and their linear equivalents. In that analysis the 
effect directly on the desorption barrier from the end groups is 
estimated to 2 kJ/mol per linear alkane for both Pt and Cu surfaces. 
We cannot judge whether the value of the cyclic-to-linear difference 
in desorption barriers \cite{lei}, 2 kJ/mol, would remain after a 
re-analysis of the desorption energies with more variation of $\nu$, 
along the lines of those of Tait et al.
 
%%%%%%%%%%%%%%%%%%%%%%%%%%%%%%%%%%%%%%%%%%%%%%%%%%%%%%%%%%%%%%%%%%%%%%%%%%%%%%%%%%%%%%%%%%%%%%
\section{Summary}
We present a computational study of the adsorption of small $n$-alkanes
on graphene using the van der Waals density functional method vdW-DF. 
Recent desorption experiments \cite{tait2006} have shown desorption 
barriers growing linearly with the size of the alkane molecule, but 
with an offset in the limit of zero length. Here we reproduce in our 
calculations the linear dependence on the alkane length including an 
offset the same size as obtained by the experiments. 
With the help of our calculated  adsorption energy of polyethylene and H$_2$ 
molecules we argue that a simple additivity assumption of alkane methylene (-CH$_2$-) units
plus two extra H atoms for the alkane ends explains the size and origin of the energy offset very well.
Summing up, our calculations thus give support to the suggestion that the 
offset measured in $n$-alkane desorption experiments
(after correction for effects of varying $\nu$) can be explained by
the $n$-alkane end groups being different from the methylene segments of the $n$-alkanes.

\acknowledgments
Partial support from the Swedish Research Council (VR) and from the
Chalmers Area of Advance Materials is gratefully acknowledged.
The computations were performed on resources provided by the Swedish 
National Infrastructure for Computing (SNIC) at C3SE. Mattias Slabanja is 
acknowledged for assistance concerning technical aspects and implementation in 
making the code run on the C3SE resources.
EKK, ML, DO, and JDR carried out their part of the research presented here
as part of the undergraduate program at Chalmers University of Technology.

%%%%%%%%%%%%%%%%%%%%%%%%%%%%%%%%%%%%%%%%%%%%%%%%%%%%%%%%%%%%%%%%%%%%%%%%%%%%%%%%%%%%%%%%%%%%%%

\end{document}